\documentclass[12pt,a4]{article}
\usepackage[english]{babel}
\usepackage{amsmath}
\usepackage{graphicx}
\usepackage{epsfig}
\usepackage{cite}
\usepackage{mcite}
\chardef\usc=95
\chardef\til=126
\catcode`\@=11 % @ signs are now treated as letters
\DeclareRobustCommand\xdotspace{\futurelet\@let@token\@xdotspace}
\def\@xdotspace{%
  \ifx\@let@token.\else
  \ifx\@let@token\bgroup.\else
  \ifx\@let@token\egroup.\else
  \ifx\@let@token\/.\else
  \ifx\@let@token\ .\else
  \ifx\@let@token~.\else
  \ifx\@let@token!.\else
  \ifx\@let@token,.\else
  \ifx\@let@token:.\else
  \ifx\@let@token;.\else
  \ifx\@let@token?.\else
  \ifx\@let@token/.\else
  \ifx\@let@token'.\else
  \ifx\@let@token).\else
  \ifx\@let@token-.\else
  \ifx\@let@token\@xobeysp.\else
  \ifx\@let@token\space.\else
  \ifx\@let@token\@sptoken.\else
   .\space
   \fi\fi\fi\fi\fi\fi\fi\fi\fi\fi\fi\fi\fi\fi\fi\fi\fi\fi}
\catcode`\@=12 % @ signs are no longer letters

\newcommand{\stru}[2]{%
   \relax\ifmmode\hbox{\vrule height#1 depth#2 width0pt}%
   \else\vrule height#1 depth#2 width0pt\fi}
\newcommand{\Ronum}[1]{\uppercase\expandafter{\romannumeral#1}}
\newcommand{\ronum}[1]{\expandafter{\romannumeral#1}}
\DeclareRobustCommand{\LaTeXZ}{%
  \LaTeX\kern-.05em4\kern-.1em
  {\raisebox{-0.2ex}{$\scriptstyle\text{ZEUS}$}}\xspace}
\DeclareMathAlphabet{\mathbf}{OT1}{cmr}{bx}{sl}
\newcommand{\eVdist}{\kern-0.06667em}

\newcommand{\gev}{{\,\text{Ge}\eVdist\text{V\/}}}

\newcommand{\slashfrac}[2]{%
  \raisebox{0.5ex}{\ensuremath #1}\kern-0.12em/\kern-0.08em
  \raisebox{-.8ex}{\ensuremath #2}}
\newcommand{\sqr}[3]{%
    {\vcenter{\hrule height.#3ex\hbox{\vrule width.#2ex height#1ex
     \kern#1ex\vrule width.#3ex}\hrule height.#2ex}}}

\catcode`\@=11 % @ signs are now treated as letters
\newcommand{\parenbar}{\mathpalette\p@renb@r}
\def\p@renb@r#1#2{\vbox{%
  \ifx#1\scriptscriptstyle \dimen@.7em\dimen@ii.2em\else
  \ifx#1\scriptstyle \dimen@.8em\dimen@ii.25em\else
  \dimen@1em\dimen@ii.4em\fi\fi \offinterlineskip
  \ialign{\hfill##\hfill\cr
    \vbox{\hrule width\dimen@ii}\cr
    \noalign{\vskip-.3ex}%
    \hbox to\dimen@{$\mathchar300\hfil\mathchar301$}\cr
    \noalign{\vskip-.3ex}%
    $#1#2$\cr}}}
\catcode`\@=12 % @ signs are no longer letters

\newcommand{\IP}{{\rm I$\kern-0.01667em$P}\xspace}

\mathchardef\qsm=63
\mathchardef\pls=43
\mathchardef\mns=512
\mathchardef\plm=518
\mathchardef\eql=61
\mathchardef\smallleft=300
\mathchardef\smallright=301
\mathchardef\les=316
\mathchardef\gre=318
\mathchardef\leq=532
\mathchardef\grq=533
\catcode`\@=11 % @ signs are now treated as letters
\newcounter{pict@width}
\newcounter{pict@height}
\newlength{\pict@scale}
\newcommand{\psfigadd}[4]{%
\setcounter{pict@width}{1*\ratio{#2+\pict@scale/2}{\pict@scale}}
\setcounter{pict@height}{1*\ratio{#3+\pict@scale/2}{\pict@scale}}
\setlength{\unitlength}{\pict@scale}
\hbox to #2{\hspace{-\fill}\begin{picture}(\thepict@width,\thepict@height)
\put(0,0){\psfig{figure=#1,width=#2,height=#3,clip=}}
\SetScale{0.283466457}
\SetWidth{1.763889}
{#4}
\end{picture}}
}
\newcounter{pict@widthfst}
\newcounter{pict@widthscd}
\newcounter{pict@widthtot}
\newcommand{\psfigaddtwo}[7]{%
\setcounter{pict@widthfst}{1*\ratio{#2+\pict@scale/2}{\pict@scale}}
\setcounter{pict@widthscd}{1*\ratio{#2+#4+\pict@scale/2}{\pict@scale}}
\setcounter{pict@widthtot}{1*\ratio{#2+#4+#6+\pict@scale/2}{\pict@scale}}
\setcounter{pict@height}{1*\ratio{#3+\pict@scale/2}{\pict@scale}}
\setlength{\unitlength}{\pict@scale}
\hbox{\hspace{-\fill}\begin{picture}(\thepict@widthtot,\thepict@height)
\put(0,0){\psfig{figure=#1,width=#2,height=#3,clip=}}
\put(\thepict@widthscd,0){\psfig{figure=#5,width=#6,height=#3,clip=}}
\SetScale{0.283466457}
\SetWidth{1.763889}
{#7}
\end{picture}}
}
\newcommand{\psfigror}[4]{%
\setcounter{pict@width}{1*\ratio{#2+\pict@scale/2}{\pict@scale}}
\setcounter{pict@height}{1*\ratio{#3+\pict@scale/2}{\pict@scale}}
\setlength{\unitlength}{\pict@scale}
\hbox{\begin{picture}(\thepict@width,\thepict@height)
\put(0,\thepict@height){\psfig{figure=#1,width=#3,height=#2,clip=,angle=270}}
\SetScale{0.283466457}
\SetWidth{1.763889}
{#4}
\end{picture}}
}
\newcommand{\psfigrol}[4]{%
\setcounter{pict@width}{1*\ratio{#2+\pict@scale/2}{\pict@scale}}
\setcounter{pict@height}{1*\ratio{#3+\pict@scale/2}{\pict@scale}}
\setlength{\unitlength}{\pict@scale}
\hbox{\begin{picture}(\thepict@width,\thepict@height)
\put(0,0){\psfig{figure=#1,width=#3,height=#2,clip=,angle=90}}
\SetScale{0.283466457}
\SetWidth{1.763889}
{#4}
\end{picture}}
}
\catcode`\@=12 % @ signs are no longer letters
\newlength\listtextwidth

\catcode`\@=11 % @ signs are now treated as letters
\newlength{\@tabfninsert}
\newlength{\@tabfnwidth}
\newcommand{\tabfootnote}[2]{%
  \setlength{\@tabfninsert}{0.8em}
  \setlength{\@tabfnwidth}{\textwidth}
  \addtolength{\@tabfnwidth}{-\@tabfninsert}
  \addtolength{\@tabfnwidth}{-0.4em}
  \noindent\makebox[\@tabfninsert][r]{\footnotesize$^{#1}$\hfil}\hfill%
  \parbox[t]{\@tabfnwidth}{\footnotesize #2\hfill}}
\catcode`\@=12 % @ signs are no longer letters

\usepackage{xspace} % to add empty space after commands

\newcommand{\PR}{Phys. Rev.\ }
\newcommand{\PRL}{Phys. Rev. Lett.\ }
\newcommand{\PL}{Phys. Lett.\ }
\newcommand{\NP}{Nucl. Phys.\ }
\newcommand{\ZP}{Z. Phys.\ }

\newcommand{\CPC}{Comp. Phys. Comm.\ }
\newcommand{\EPJ}{Eur. Phys. J.\ }

\newcommand{\beq}{\begin{equation}}
\newcommand{\eeq}{\end{equation}  }

\usepackage{color,fancybox,amsfonts,babel,graphics,graphicx}
\makeatother
\begin{document}

\hfill{}LC-PHSM-2003-001

\hfill{}ANL-HEP-PR-03-002

\hfill{}January 7, 2003

\vspace{2.0cm}

\bigskip{}
\begin{center}
\textbf{\large Selection and reconstruction of the top
quarks in the all-hadronic decays 
at a Linear Collider}
\end{center}{\large \par}

\vspace*{1.0cm}

\begin{center}S.~V.~Chekanov $^a$ and V.~L.~Morgunov $^b$
\end{center}

{\begin{itemize}
\itemsep=-1mm
% \leftskip-0.5cm
\normalsize
\item[$^a$]

\small
HEP Division, Argonne National Laboratory,
9700 S.Cass Avenue,
Argonne, \\ IL 60439 USA

\normalsize
\item[$^b$]

\small
DESY, Notkestrasse 85, D-22603 Hamburg, Germany and  \\  
ITEP, B.Cheremushkinskaya, 25, Moscow, 117218, Russia
\end{itemize}
}

\normalsize
\vspace{1.0cm}

%%%%%%%%%%%%%%%%%%%%%%%%%%%%%%%%%%%%%%%%%%%%%%%%%%%%
\bigskip{}

\begin{abstract}
A method of reconstruction of the top quarks produced in the
process $e^{+}e^{-}\to t\overline{t}\to 6\, {jets}$ at 
a Linear Collider (LC) is proposed. The approach does not involve 
a kinematic fit, as well as 
assumptions  on the invariant masses of the dijets originating from 
the decays of $W$ bosons and, therefore, 
the method  is expected to be less sensitive to 
theoretical and experimental
uncertainties on the top-mass measurement 
than traditional reconstruction methods. For the first time,
the reconstruction of the top quarks was investigated  
using the full LC detector
simulation after taking into account the background arising 
from QCD multi-jet production.
\end{abstract}

%%%%%%%%%%%%%%%%%%%%%%%%%%
\section{Introduction}
%%%%%%%%%%%%%%%%%%%%%%%%%%

The standard procedure of the top-quark measurements in the process
$e^{+}e^{-}\to 
t\overline{t}\to 
b\overline{b}q_{1}\overline{q}_{2}q_{3}\overline{q}_{4}\to 6\, {jets}$
takes advantage of a fully reconstructed final state. 
The decay signature is characterized by the production of 
six hadronic jets,
therefore, a background arising from the standard 
six-jet QCD processes is expected. 
Nevertheless, since the all-hadronic top decay  
has the largest branching ratio ($\simeq 44\%$ of all 
$t\overline{t}$ decays),  
it is considered as one of the most promising at the 
TEVATRON, LHC and a Linear Collider (LC).  

A reduction of the background from QCD multi-jet events
can be achieved by identification of $b$-initialized jets (the
so-called $b-$tagging). Furthermore, one can use  
the requirement that the invariant mass of two jets 
(not associated with the $b$ quarks) is consistent 
with the decay of the $W$ bosons.
These two methods were used for the
top reconstructions at the TEVATRON (see, for example,
\cite{tevat}), and also are considered for the LHC \cite{lhc}
and  LC experiments 
\cite{iwasaki,epj:c9:229,*np:b544:289,*gangemi,che:lcsim}. 

The reconstruction of the top quarks in the all-hadronic decays
is known to be affected by many theoretical 
uncertainties:
the extraction of the top-quark pole
mass has a theoretical uncertainty of around 300 MeV and cannot
be determined with a precision better than
${\cal O} (\Lambda_{QCD})$ 
\cite{pl:b434:115,*pr:d59:114014,*EPJdirect:c3:1}. 
In addition, incomplete knowledge of the hadronic final
state leads to an uncertainty of a 
few hundreds MeV \cite{che:lcsim}.  For the latter,
one of the most significant uncertainties is related to the
reconstruction of the $W$ mass, $M_W$, 
after minimizing $\mid M_{jj} - M_W \mid$, where  $M_{jj}$ is
the dijet invariant mass.   
The way how the large Breit-Wigner tails of the 
decay $W\to 2\>\> jets$ 
is treated when the top quarks are reconstructed 
directly affects the reconstructed top mass \cite{che:lcsim}. 
A significant impact on the reconstructed top mass 
may come from non-perturbative phenomena, such as 
colour rearrangement 
\cite{zp:c62:281,*epj:c6:271,*zp:c72:71,*epjdir:c1:1,*kh_sj_top} 
and Bose-Einstein effects \cite{pl:B351:293,*epj:c6:403},  
which are expected to shift the reconstructed masses 
in the $W^+W^-$ and $t\bar{t}$ production. 
Thus, any approach involving the  
reconstruction of $W\to 2\>\> jets$ is
bound to lead to a systematic shift for the top reconstruction.
For example, it has been shown that the present implementation
of the Bose-Einstein effect in a Monte Carlo model can contribute to a
shift
in the reconstructed top mass through a distortion of the $M_{jj}$ spectrum 
\cite{che:lcsim}. Similar effects are expected for the color-reconnection
phenomenon which is anticipated to be significant for the LC energies.

In this article, a simple reconstruction method which does not involve
the direct measurement of the decay $W\to 2\>\> jets$ is proposed. 
This method is a very general, and is
suitable for the reconstruction of any process in which a particle ($V$) and 
an anti-particle $\bar{V}$ are produced and then decay as:
$$
e^+e^- \longrightarrow  V + \bar{V}  \longrightarrow
v_a V_a +  v_b V_b
$$
$$
V_a  \longrightarrow  V_1 + \bar{V}_2, \qquad 
V_b  \longrightarrow  V_3 + \bar{V}_4, 
$$
without {\em a prior} knowledge on the masses of the intermediate
particles, denoted as $V_a$  and $V_b$. 
The effectiveness of this
method is based on the assumptions that the initial-state particles,
$V$ and $\bar{V}$,  have similar (but not equal!)  masses, 
$\mid M_{V}- M_{\bar{V}} \mid << M_{V(\bar{V})}$, and that  
the simplicity of $e^+e^-$ annihilation allows to use the momentum conservation,
$\vec{P}_{V}+\vec{P}_{\bar{V}}=0$.  
These two requirements, together with the fact that both initial 
particles decay similarly 
into three other particles, are essential in the reconstruction 
of the invariant masses
of $V$  and $\bar{V}$ through  six jets in the final state.
 
%%%%%%%%%%%%%%%%%%%%%%%%%%%%%%%%%%
\section{Reconstruction procedure}
%%%%%%%%%%%%%%%%%%%%%%%%%%%%%%%%%%%

We consider $e^{+}e^{-}$ collisions in the laboratory frame. As a
first step, six jets in the event have to be reconstructed. We use the
$k_{\bot }$ (Durham) algorithm \cite{pl:b269:432}, requiring exactly six jets
in every event, without any specific cuts on the transverse momenta and
rapidity of the jets. 

%%%%%%%%%%%%%%%%%%%%%%%%%%%%%%%%%%
\subsection{Event selection}
%%%%%%%%%%%%%%%%%%%%%%%%%%%%%%%%%%%

To preselect hadronic events, we use the typical LEP cuts which take
advantage of  the energy-momentum conservation requirements:
\begin{equation}
\mid \frac{E_{vis}}{\sqrt{s}}-1\mid <\Delta_E,
\quad \frac{\mid \sum \vec{p}_{||i}\mid }{\sum \mid \overrightarrow{p_{i}}\mid }<\Delta_{PL},
\quad \frac{\mid \sum  \vec{p}_{Ti}  \mid }
{\sum \mid \overrightarrow{p_{i}}\mid }<\Delta_{PT} ,
\label{eq:1}
\end{equation}
where $E_{vis}$ is the visible energy, $\vec{p}_{||i}$ ($\vec{p}_{Ti}$) is
the longitudinal (transverse) component of momentum of a final-state particle.
The $\Delta_E$, $\Delta_{PL}$ and $\Delta_{PT}$ are small adjustable parameters.
The top events are characterized by a large amount of missing energy/momentum due to
undetected neutrinos, therefore, the selection (\ref{eq:1}) is especially
important as it helps to reject events with a 
significant fraction of neutrinos \cite{che:lcsim}.

The jets in the all-hadronic top-decay channel should be well separated in the
transverse momenta.  
The best way to do this is to apply a restriction on the values of
$y^{cut}_6$  used in the Durham algorithm to resolve  six jets. 
We accept only events if $y^{cut}_6>\Delta_y$, with  $\Delta_y$ being 
a parameter to be found
using a Monte Carlo simulation. 
A similar cut was applied for the JADE jet algorithm in
the study of the all-hadronic top decay using the LCD Fast Simulation \cite{iwasaki}.

%%%%%%%%%%%%%%%%%%%%%%%%%%%%%%%%%%
\subsection{Top reconstruction}

%%%%%%%%%%%%%%%%%%%%%%%%%%%%%%%%%%%

We start with the initial list of six jets with momenta $p_{i}$. These
jets are merged into groups, with three jets in each group. 
Thus, every event can be considered to consist of pairs of the 
three-jet groups with momenta $\{P_{I}(1),\, P_{I}(2)\}$, where 
$P_{I}(1)\equiv P_{i,j,m}=p_{i}+p_{j}+p_{m}$ is the four-momentum of each group,
and $P_{I}(2)$ was obtained analogously using the rest of the jets
in the same event. In total, there are twenty three-jet groups, which
can be arranged into ten three-jet pairs with non-identical
momenta, i.e. $I=1,\ldots 10$.
Ideally, from these ten pairs of three-jet clusters, one should accept only
one three-jet pair corresponding to two initially produced  particles. 
In practice, 
it is difficult to find the correct assignments of jets due to 
fragmentation and detector effects. Therefore,
a special selection should be performed in order to 
minimize the number of three-jet pairs. Assuming that the produced
particles have similar masses,
we accept such three-jet pairs if the 
invariant masses of the three-jet groups satisfy the condition
$|M_{I}(1)-M_{I}(2)|< \Delta_M$, where
$M_{I}$ is the invariant mass of  a three-jet group with the momentum
$P_{I}$, and $\Delta_M$ is a free parameter constraining the 
invariant masses of three jets inside each pair.

Next, energy-momentum conservation can be used to set an additional 
constraint  on the momenta of three-jet groups within each three-jet pair. 
We accept only three-jet groups which are produced back-to-back, 
i.e. we require
$\mid \overrightarrow{P}_{I}(1)+\overrightarrow{P_{I}}(2) 
\mid< \Delta_P$, where
$\Delta_P$ is again an adjustable  parameter.

\medskip

The following additional selection criteria may also be used:
\begin{itemize}
\item 
The list of three-jet groups  contains top-quark candidates, which
decay as $t\to bW\to bq\overline{q}$. In order to reduce the number
of possible three-jet pairs, one can use an efficient double $b$-tagging,
requiring one and only one $b$-initiated jet within each three-jet group.
All three-jet pairs which have at least one three-jet group 
with two $b$-initiated jets have to be removed.
This leads to six possible three-jet pairs, compared to ten if
no $b$-tagging is used.
 
\item
Each three-jet group should contain two jets coming from the 
decay of the $W$ boson.
Thus, one can remove such three-jet groups which cannot
be associated with the $W$-decay hypothesis. This can be done
by accepting such three-jet pairs which have the invariant masses of any two
jets inside the region $\mid M_{jj} -M_{W}\mid < \Delta_W$. 
This requirement is not advocated in this article
as it makes the specific physics assumptions on the
invariant mass and width of the $W$ boson. 

\end{itemize} 

These two additional criteria are optional; in fact, 
we will show that even without them
one can obtain a robust reconstruction procedure.

%%%%%%%%%%%%%%%%%%%%%%%%%%%%%%%%%%%%%%
\subsection{Monte Carlo studies}
\label{sec:gen}
%%%%%%%%%%%%%%%%%%%%%%%%%%%%%%%%%%%%%%

To illustrate the method outlined above, we first apply it to the
all-hadronic top decays in $e^+e^-$ annihilations
at the centre-of-mass energy of $\sqrt{s}=500$ GeV 
generated using the PYTHIA 6.2 model \cite{cpc:82:7}. 
The contribution from the initial-state radiation (ISR) 
can be rather significant at 
LC energies, thus this effect was included in the simulation.
The mass and the Breit-Wigner width of the top quarks were set to 
the defaults values, 175 GeV and 1.39 GeV, respectively. 
The particles with the lifetime more than 3 cm are
considered to be stable. Neutrinos were removed from the consideration.

\begin{figure}[t]
\begin{center}
\includegraphics[  scale=0.65]{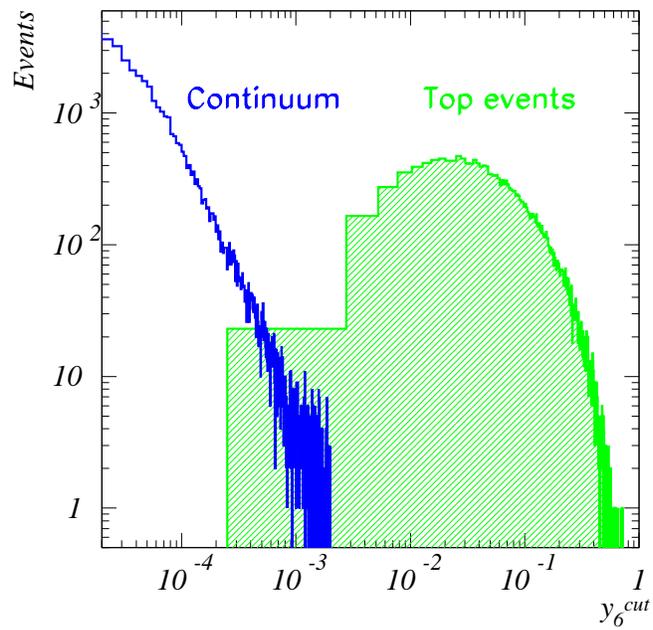}
\caption{The values of $y^{cut}_6$ for the all-hadronic top decays 
and for the rest of inclusive $e^+e^-$ sample (continuum) generated with the
PYTHIA model.} 
\label{cap:y6_cut}
\end{center} 
\end{figure}

In order to reconstruct the top quarks using the proposed method, one
should define the following: a) the parameters for the event selection: 
$\Delta_E$, $\Delta_{PL}$, $\Delta_{PT}$, $\Delta_y$; b) 
the parameters for the  final reconstruction: $\Delta_M$ and $\Delta_P$. 
The event-selection parameters should not be large in order to 
insure small  contaminations from neutrinos resulting to a broad and 
asymmetric Breit-Wigner peak for the three-jet mass distribution. 
In this study, we will apply rather tight cuts: 
$\Delta_E = \Delta_{PL} =\Delta_{PT} =0.02$. 
The parameter $\Delta_y$ requires an additional study: 
Fig.~\ref{cap:y6_cut} shows the distribution for the all-hadronic top decays and 
for $e^+e^-$ inclusive  events 
(but without the $e^+e^-\to t \bar{t}\to 6\> jets$ process) 
generated with the PYTHIA model. 
The top events are characterized by $y^{cut}_6 > \Delta_y = 2\cdot 10^{-4}$, 
thus this value is used for the studies below.

The parameter $\Delta_M$ controls the maximum difference between masses of 
two top-quark candidates.
It was set to 30 GeV, which is large enough to insure
a reasonable event acceptance, but sufficiently   small 
to exclude wrong assignments of jets. 
The parameter $\Delta_P=5$ GeV
further reduces the contributions from unmeasured neutrinos.
Note that the latter cut is not very important for the 
generator-level study, since the event selection
is already very strong. In practice, such cuts
should not be very tight to obtain a reasonable acceptance after a 
detector reconstruction. 

\begin{figure}
\begin{center}
\includegraphics[  scale=0.65]{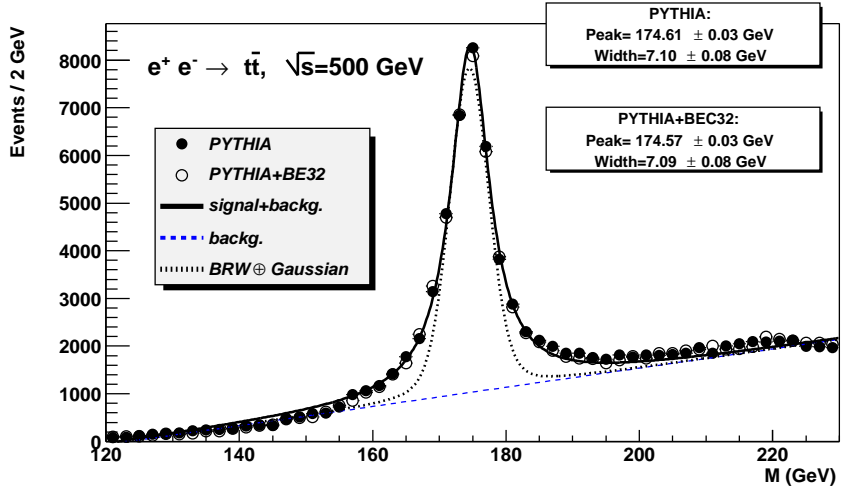}
\caption{The distribution of three-jet invariant masses for
$e^+e^-\to t \bar{t} \to 6 jets$ events at $\sqrt{s}=500$ GeV.
The all-hadronic decay channel was generated with the PYTHIA model with (open dots) 
and without (solid dots) the Bose-Einstein effect. The peak position and the 
width were determined from 
the Breit-Wigner distribution, which gives the best $\chi^2$ for the fit.
A Gaussian  convoluted with the Breit-Wigner distribution
(with the fixed width of 1.39 GeV) is also shown (dashed line).   
The first-order polynomial is used to describe the background.
The solid line shows the fit for the invariant mass obtained from
the PYTHIA models without the BEC32 effect included.
}
\label{cap:bfig1a}
\end{center}
\end{figure}

Figure~\ref{cap:bfig1a} shows the three-jet invariant masses  after the
selection procedure described above, but without the requirements 
of double $b$-tagging, and also without the cut on 
$\mid M_{jj}-M_W\mid$.  Two different
functions were used to fit the peak: 
a) the Breit-Wigner distribution (with the fixed 
width of 1.39 GeV) convoluted with a Gaussian and  
b) the Breit-Wigner function alone. 
The first-order polynomial is used in order 
to describe the background in  both cases.
The motivation for the Gaussian convoluted with the Breit-Wigner function
comes from the assumption that contributions of many
independent effects (parton-shower splittings, 
hadronization and resonance decays) should yield a 
Gaussian-like distribution.
As seen from Fig.~\ref{cap:bfig1a}, the best fit
can be obtained by using the Breit-Wigner function alone, while
the Gaussian does not describe adequately the tails of the
mass distribution.
Based on this empirical observation, we will use the Breit-Wigner
function to fit the mass spectra before the LC detector simulation.

The fact that the Gaussian distribution fails
to describe the mass spectrum might indicate  
a strong correlation 
between different effects
leading to non-Gaussian broadening of the natural width.
It was verified that a sum of two Gaussian
distributions gives a better fit than 
the Gaussian distribution alone, but the Breit-Wigner still gives
the best fit.
It is likely that a sum of many Gaussian distributions should be used
to fit the mass spectrum. 
In this case, an unknown weight for each Gaussian contribution  
would reflect the fraction of events with a particular property.

One interesting feature of the invariant-mass distribution 
shown in Fig.~\ref{cap:bfig1a}
is that it has no background in the region of small three-jet  
masses, where the major contribution from the QCD multi-jet background
is expected. This can be explained as following: any misassignment of jets at a
centre-of-mass energy far from the top-production threshold usually
increases the observed three-jet invariant mass, since in this
case the misassignment jet usually has a significant angle with respect
to the rest of the jets in the three-jet group. This ultimately increases the
three-jet invariant masses.

As was mentioned in the introduction,
this method is expected to be less sensitive to the hadronic-final
state uncertainties which usually arise when 
the invariant mass of the 
dijets from the decay $W\to 2\> jets$ is reconstructed.
Fig.~\ref{cap:bfig1a} shows the three-jet invariant-mass distributions 
without and with the Bose-Einstein interference 
included. We use the so-called ``BE32'' type of the Bose-Einstein
simulation, which is included in the PYTHIA 6.2 model \cite{cpc:82:7}.
It is clear that no any significant shift for the reconstructed top
mass can be attributed
to the Bose-Einstein effect when the proposed method is used for the
top reconstruction.

%%%%%%%%%%%%%%%%%%%%%%%%%%%%%%%%%%%%%%%%%%%%%%%%%%
\section{Top reconstruction using the full \\  detector simulation}
\label{sec:dr}
%%%%%%%%%%%%%%%%%%%%%%%%%%%%%%%%%%%%%%%%%%%%%%%%%%

The reconstruction of the top quarks after the full detector simulation was performed 
in a few steps:

1) Fully inclusive $e^+e^-$ annihilation events 
were generated with the PYTHIA 6.2 model, which was based on the
default parameters. Events with the $W^+W^-Z$ and $ZZZ$ production
leading to six jets in the final state were not included in the
simulation, as they have a small contribution ($\sim 2\% -4\%$). 
In total, 270k events were generated. This sample approximately 
corresponds to twenty days of the LC data taking.  

2) From this sample, $e^+e^-\to t\bar{t}\to {\mathrm everything}$ 
events were passed through the full GEANT-based detector simulations.
The number of such events was 9.6k. 
The TESLA detector \cite{tdr-tesla} simulation based on 
the  BRAHMS 305 \cite{brahms} program was used 
for the event reconstruction. 
Of particular importance for  the
present study  are the tracking system based on the 
Time Projection Chamber (TPC) and the calorimeter (CAL).
The TPC is  surrounded by the CAL, which
is longitudinally segmented into electromagnetic
and hadronic sections. The electromagnetic calorimeter is based on tungsten absorber and 
silicon diode pads, while the hadronic part is an Fe/scintillating tile calorimeter.
All particles in events were reconstructed using a combination of track and calorimeter
information that optimizes the resolution of the 
reconstructed jets \cite{morgunov:2001cd}.
 
3) The rest of the sample (without the top quarks) 
was treated  differently: 
Energies of the final-state particles in such events were 
smeared around the true values according to the 
Gaussian resolution functions in order to imitate the detector
response. 
Charged hadrons  (with energy denoted by $E_{ch}$) were smeared 
using the Gaussian distribution with $\sigma = 10^{-4}E^2_{ch}$.
Analogously, photon energies ($E_{\gamma}$) were smeared using
$\sigma=0.12\sqrt{E_\gamma}$, while the energies ($E_n$) 
of neutral hadron were distributed according to $\sigma=0.4\sqrt{E_n}$.
Particles close to the beam pipe were removed, 
to reproduce the geometrical acceptance of the LC detector.  
 
\begin{figure}
\begin{center}
\includegraphics[  scale=0.55]{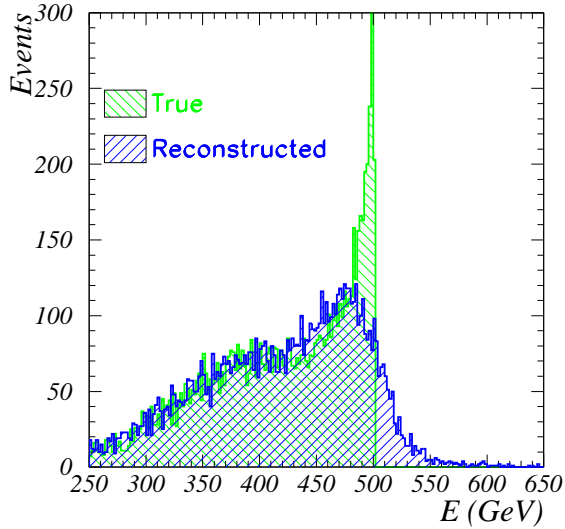}
\caption{The distributions of the total event 
energies of $e^+e^-\to t\bar{t}$ 
events before and after the LC detector reconstruction.}
\label{cap:enr_distr}
\end{center}
\end{figure}

\begin{figure}
\begin{center}
\includegraphics[  scale=0.55]{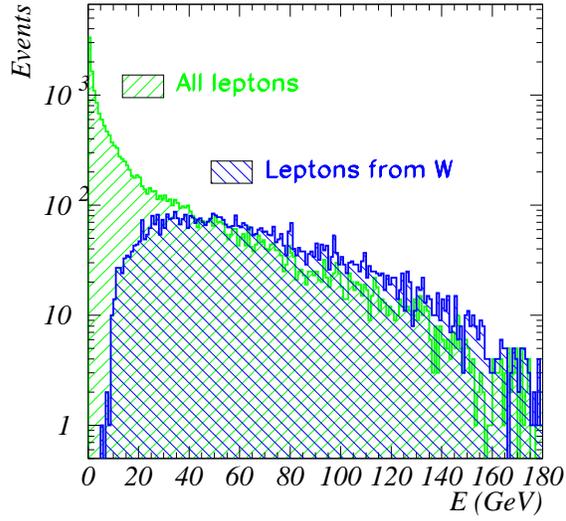}
\caption{The reconstructed energies of the final-state leptons for semileptonic 
top decays and for the $e^+e^-\to everything$ events.}
\label{cap:lepton}
\end{center}
\end{figure}

The selection and the reconstruction of top quarks at $\sqrt{s}=500$ GeV 
are  based on the parameters given in
Table~\ref{tab1}a). These parameters are the same for the generated  and
reconstructed event samples.
For the former sample, neutrinos were removed, as well as  particles which are
inside the beam pipe.  
The cuts on the energy/momentum imbalance were determined after the study of the detector
resolution, and were optimized to obtain a reasonable efficiency
of the reconstruction.
Figure~\ref{cap:enr_distr} illustrates the distributions of the total event 
energy before and after the detector simulation. A significant
missing energy due to neutrinos and the ISR 
is observed even before the event reconstruction. 
The event-selection cuts are by a factor two-three larger
than the resolution on the corresponding variable. 
Events were removed if at least one lepton is found with an energy above
$20$ GeV;  the motivation for this cut is clear from Fig.~\ref{cap:lepton}. 

\begin{figure}
\begin{center}
\includegraphics[  scale=0.5]{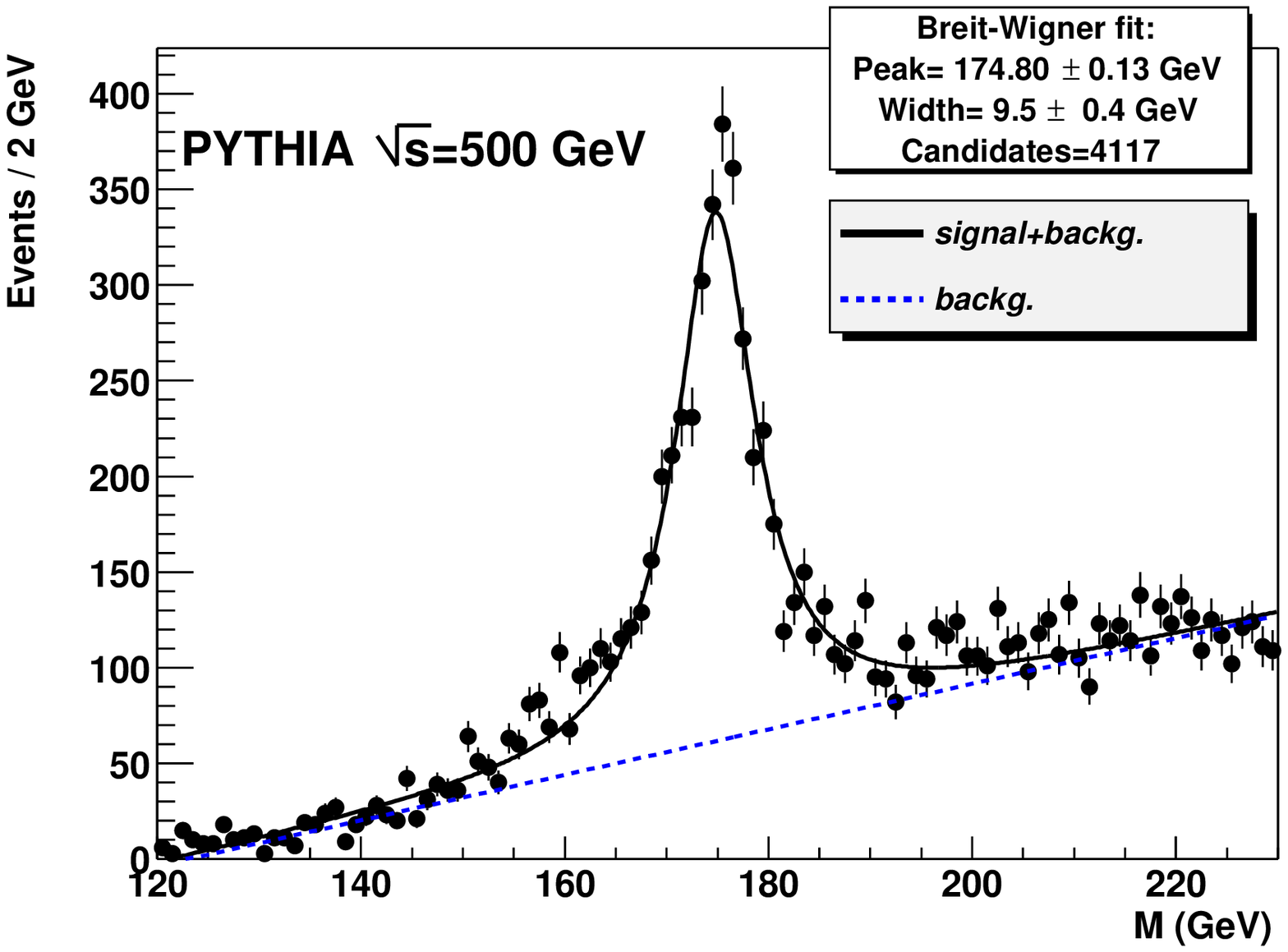}
\caption{
The invariant-mass distributions for three-jet clusters 
for $e^+e^- \to t\bar{t}$ process at $\sqrt{s}=500$ GeV 
generated using the PYTHIA model before the LC detector simulation.
The all-hadronic decays were
selected and reconstructed using the parameters given in Table~\ref{tab1}a).
} 
\label{cap:tru1}
\end{center}
\end{figure}

Figure~\ref{cap:tru1} shows the three-jet invariant masses for
the reconstructed all-hadronic decays in the process $e^+e^- \to t\bar{t}$ 
before the detector simulation.  
The width of the Breit-Wigner distribution, 9.5  GeV,
is somewhat larger than that shown in Fig.~\ref{cap:bfig1a},
since now a larger contribution from the missing energy/momentum is allowed to
obtain a higher selection efficiency. This width will be used in the
following figures for the Breit-Wigner function. 

Figure~\ref{cap:sim1} shows the three-jet invariant masses after the
full detector simulation and event reconstruction.
The  background events from the non-top continuum were added  
after the Gaussian smearing. 
The cuts are the same as for Fig.~\ref{cap:tru1}, except for the cut
on $y_6^{cut}$. 
A significant reduction of the background can be obtained by selecting
events with $y_6^{cut}> 2\cdot 10^{-4}$, as seen from Fig.~\ref{cap:sim2}.
Note that this cut is the same as for the generator-level 
studies, since the LC detector simulation does not distort significantly 
the $y_6^{cut}$ distribution.

Additional assumptions can further increase the 
signal-over-background ratio.
For example, Fig.~\ref{cap:sim3} shows 
that the  cut on the invariant 
mass of two jets inside each three-jet group, $\mid M_{jj}-M_W\mid < 15 $ GeV,
reduces the background. 
Further, a check using the generator-level reconstruction 
indicated that an  efficient double $b$-tagging
can decrease the background by $\sim 50\%$. This, however, has no 
a significant impact on the Gaussian width, which is $5.0-5.5$ GeV 
for the the current detector design and event reconstruction.

%%%%%%%%%%%%%%%%%%%%%%
\section{Top reconstruction at $\sqrt{s}=800$ GeV}
%%%%%%%%%%%%%%%%%%%%%%

The proposed method is well suited for a higher 
centre-of-mass energy of $e^+e^-$ collisions. 
The reconstruction is expected to be even more 
reliable: 
Since the hadronic jets are better collimated along 
the momenta of the produced top quarks, 
any wrong jet assignment significantly
increases the invariant mass for such a  
miss-reconstructed top candidate. This should lead
to a large invariant-mass difference between  miss-reconstructed
three-jet groups, therefore, such groups which will be removed by the cut on 
$|M_{I}(1)-M_{I}(2)|$ more effectively.   

At higher energies, however, missing event energy/momentum  
is larger due
to higher energies of neutrinos. Therefore, the selection cuts should
further be optimized to obtain a reasonable selection efficiency. 
Table~\ref{tab1}b) shows the event-selection and top-reconstruction cuts
applied for  $\sqrt{s}=800$ GeV. 
We also use  the same reconstruction
cuts as for $\sqrt{s}=500$ GeV to be able to compare the results with 
the previous founding. 
Note that the $y_6^{cut}$ distribution for $\sqrt{s}=800$ GeV 
is somewhat different than that
shown in Fig.~\ref{cap:y6_cut}: now the distribution extends down to
$y_6^{cut}=10^{-5}$, and has a small peak at 
$y_6^{cut}\sim 10^{-4}$. This is due to a better
collimation of jets originating from the top quark. It was found,
however,  that a decrease of  $\Delta_y$ for events at $\sqrt{s}=800$ GeV
has a small impact on the final selection efficiency.

\begin{figure}
\begin{center}
\includegraphics[  scale=0.5]{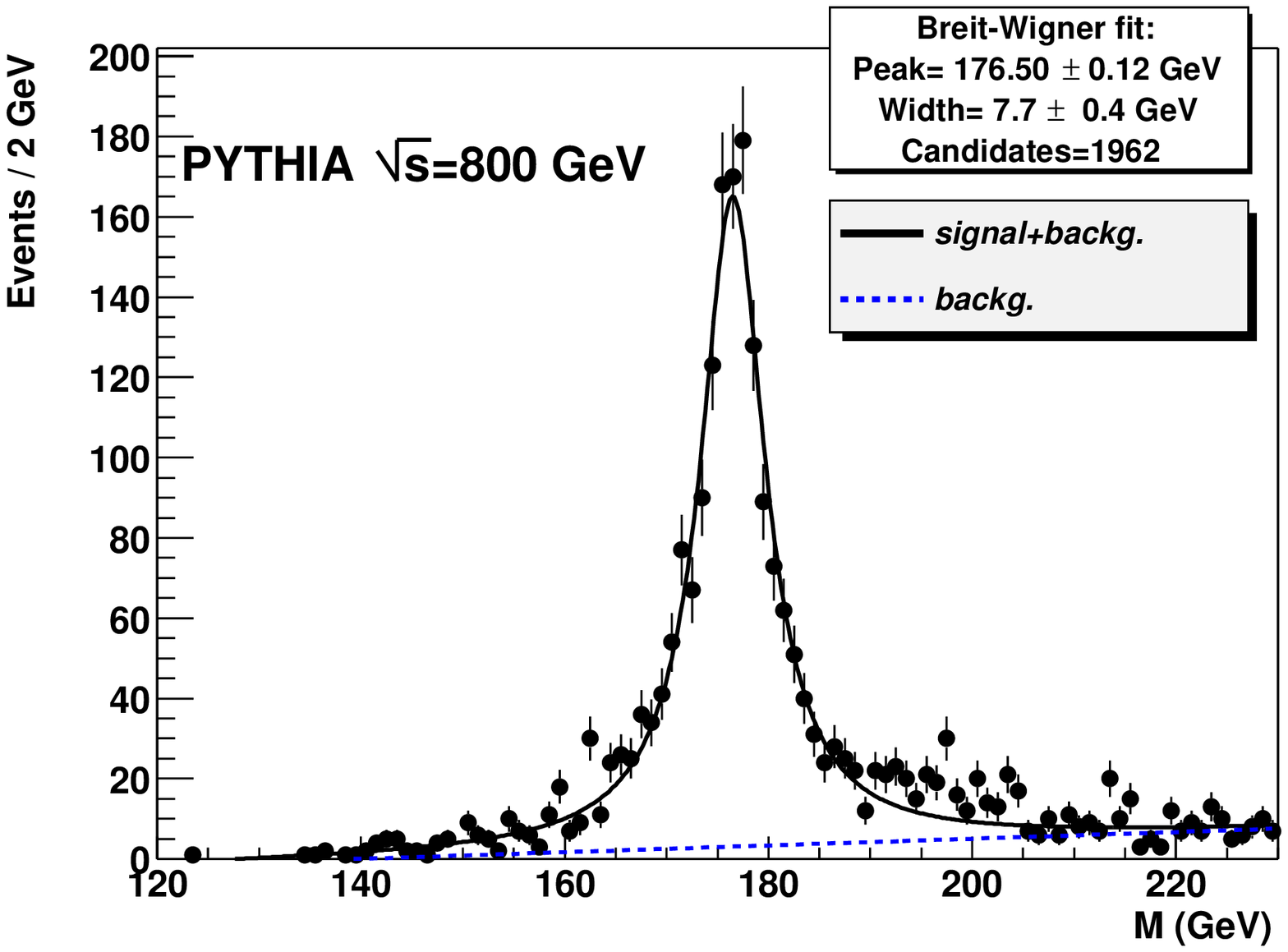}
\caption{
The invariant-mass distributions for three-jet clusters 
for $e^+e^- \to t\bar{t}$ process at $\sqrt{s}=800$ GeV generated using the PYTHIA model.
The all-hadronic decays were
selected and reconstructed using the parameters given in Table~\ref{tab1}b).
} 
\label{cap:tru2}
\end{center}
\end{figure}

Figure~\ref{cap:tru2} shows the three-jet invariant masses 
for $e^+e^-\to t\bar{t}$ events generated with PYTHIA before the detector
simulation. The PYTHIA event sample has the same size as for Sect.~\ref{sec:dr}.
The signal-over-background ratio
is significantly larger for events at $\sqrt{s}=800$ GeV 
than for $\sqrt{s}=500$, as expected.

Note that the efficiency of the top selection
is less by a factor two than for $e^+e^-$ annihilation at
$\sqrt{s}=500$ GeV.
It was found that most of the events were rejected by the cut 
$|\overrightarrow{P}_{I}(1)+\overrightarrow{P_{I}}(2)|< \Delta_P$, 
thus this cut is the main  in reducing 
the fraction of events with significant missing energies from neutrinos.
A check without the cut on the energy/momentum
imbalance shows a very similar result 
as for the case without these cuts included. 
Generally speaking, the cuts on the event imbalance can be applied mainly 
to reduce the computational time before the top reconstruction, 
rather than to improve the signal-over-background ratio.  

Figure~\ref{cap:sim5} shows the reconstructed signal after the
detector simulation and when events from the continuum were included.
No any cuts on the dijet invariant mass were applied, i.e. this figure
is equivalent to Fig.~\ref{cap:sim2} for the 
lower centre-of-mass energy. The signal for
$\sqrt{s}=800$ GeV has significantly less background, however, the number
of the reconstructed top candidates
is also smaller due rejection of events with large energy losses.  

It is important to notice that the width of the Gaussian distribution is larger for
$\sqrt{s}=800$ GeV than for $\sqrt{s}=500$ GeV. There are a few
reasons for this. First, the energy resolution of the CAL is worse for
higher jet energies. This, however, cannot completely explain the observed
increase in the mass width. Again, let us remind that 
jets from the decay of a single top are better
collimated at $\sqrt{s}=800$ GeV, thus 
a larger degree of overlaps between energy deposits in the
calorimeter is expected that makes more difficult to reconstruct
the energy-flow  objects. Furthermore, one should expect a stronger
leakage for neural particles outside the calorimeter.   

%%%%%%%%%%%%%%%%%%%%%%%%%%%%%%%%%%%%%%%%%%%%%%%%%
\section{Snagging  the top quarks using a neural network}
%%%%%%%%%%%%%%%%%%%%%%%%%%%%%%%%%%%%%%%%%%%%%%%%%%

Neural networks (NN) have seen an explosion of interest over the last
years, and have been successfully applied for paten recognition problems
in particle physics. Based on the top-reconstruction
approach described above, in this section we will devise a NN method to select
top events using the hadronic-final-state signatures of the all-hadronic decays.

\begin{figure}
\includegraphics[scale=0.8]{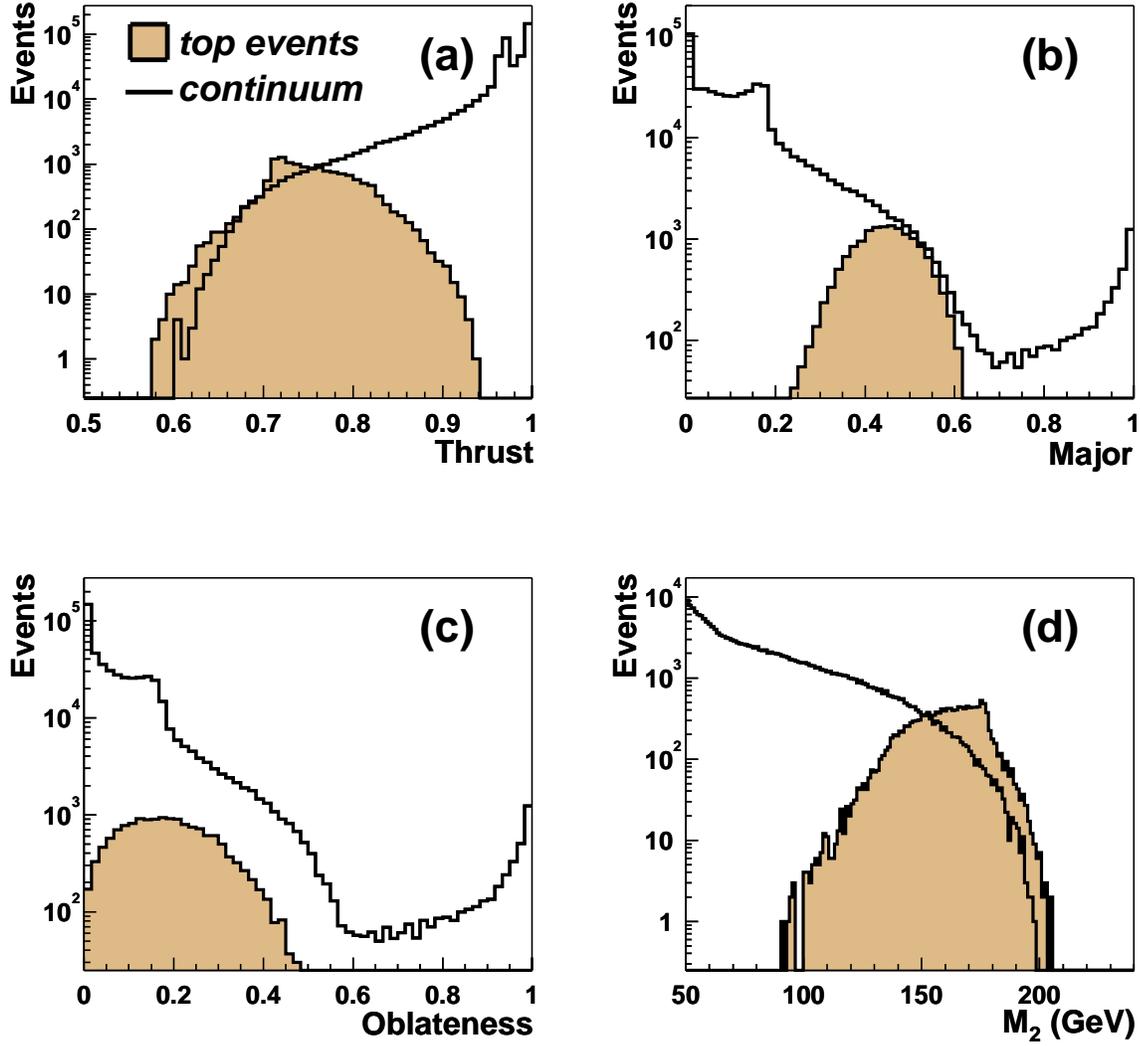}
\caption{The values of the thrust, major, oblateness and 
the invariant masses of event hemispheres  
for fully inclusive $e^{+}e^{-}$ PYTHIA events
generated  without the $t\bar{t}\to\> 6 jets$ process 
and for the all-hadronic top decays (shaded aria).} 
\label{cap:shapes}
\end{figure}

As a first step, one should define the variables for NN input
which are likely to be influential. The event-shape variables represent
the most efficient way to separate the $t\bar{t}$ from the 
continuum \cite{igo-1992,epj:c9:229}. The thrust ($T$) 
\cite{pl:12:54,*prl:39:1587}, major ($Mj)$
and oblateness $(Ob)$ event characteristics will be 
our primary choice. The allowed
range for thrust is $0.5\leq T\leq 1$, such that the isotropic events
have $T\sim 0.5$, whilst a 2-jet configuration should have $T\sim 1$.
We use as positive direction of the thrust axis the direction of the
most energetic jet. The thrust axis is defined as the axis along which
the projected energy flow is maximized. The major is sensitive to
the energy flow in the plane perpendicular to the event thrust axis,
and its direction is defined in the same fashion as thrust. The minor
axis is the third axis, which is perpendicular to the thrust and major
axis. The difference between the major and minor values is called
the oblateness, such that $Ob\sim 0$ corresponds to an event configuration
symmetrical around the thrust axis. 
Figs.~\ref{cap:shapes}a)-c) 
show the values of the thrust, major and oblateness for the PYTHIA
6.2 model, after the energy smearing
to imitate the LC detector response. The shaded
band represents the $e^+e^-\to t\bar{t} \to 6\> jets$ events, 
scaled according to the predicted cross section.
It is clear that the values of
the thrust and the major have the best sensitivity the top-quark production,
while the  oblateness reflects the properties of the top decays in less extent.

It is mandatory to use an additional information on the invariant mass
of the decaying system, together with  the variables discussed above
which mainly focus on the topological properties of the event shapes.
Since the top mass is approximately known, and because the production
of two top quarks at $\sqrt{s}\geq 500$ GeV 
is above the top-production energy threshold, an  
$e^+e^- \to t\overline{t}$ event
can be viewed as a simple two-body decay with non-overlapping decay
products. The top events can be divided in two event hemispheres,
each of which has an invariant mass close to the top mass. 
The hemispheres can be defined by using the thrust axis, such
that particles with positive $Z$ component belong the first hemisphere,
while particles with $Z<0$ form the second hemisphere. Fig.~\ref{cap:shapes}d)
shows the invariant-mass distribution, $M_2=0.5\, (M_{a}+M_{b})$, where
$M_{a}$ ($M_{b}$) is the invariant mass of the hadronic system in
the first (second) hemisphere. For the $e^+e^-\to t\bar{t} \to 6\> jets$ events, 
the invariant mass
has a broad peak near the generated (true) top-mass value, while 
$e^+e^-$ events without the all-hadronic top decays 
typically have rather low invariant masses.

Furthermore, 
since the all-hadronic top 
decay is  characterized by 
a six-jet configuration, one can use  this information 
for the top selection. Particularly, $y^{cut}_N$ values 
of the resolution parameter of the 
Durham algorithm for which an $N$-jet configuration becomes 
an $N+1$ jet system can also be useful.
We have already illustrated that the cut on
$y^{cut}_6$ alone has a significant impact on the selection of the all-hadronic top decays. 

\subsection{Standardizing inputs}

The contribution of an input for a NN depends heavily on its variability
relative to other inputs. Therefore, it is essential to rescale the inputs,
so their variability reflects their importance. We will rescale some
inputs to the interval $[0,1]$, which is the most popular 
choice for the NN inputs.

In case of the event shape distributions, we do not use any rescaling,
since all these variable are already in the interval $[0,1]$. For 
the mass distribution $M_2$, we used the function
$A_{1}(\ln M_2 -A_{2})$, with $A_{1}=3.4758$ and $A_{2}=5.0106$.
This choice was advocated in \cite{pl:b278:181}, and here we only adjusted
the free parameters $A_{1}$ and $A_{2}$, such that 
this new variable is zero for $m_{t}=150$ GeV, and 
it equals to unity when $m_{t}=200$ GeV.

The situation with $y^{cut}_N$ values is more complicated, since their
values strongly depend on $N$. As example, $y_{2}^{cut}$
for which 2 jets become a 3-jet system has a significantly larger
value than $y_{N>2}^{cut}$.
For the all-hadronic top decay, the relative importance of $y_{N}^{cut}$
values for event configurations with large $N$ is higher than for low
$N$. Therefore, in order to
treat all inputs on an equal footing, we transformed all $y_{N}^{cut}$ values
according to the expression $Y_{i}=i^{2}\, y_{i+1}^{cut}\, \, i=1\ldots 6$,
which increases the relative weights of $y^{cut}_N$ values in the NN
input for events with many jets.

\subsection{Training sets}

We have trained  the NN to learn just one number in the range between
0 and 1. When the NN output is unity, this corresponds to the most
probable top candidate in the all-hadronic decays.

A few data sets were used. The first event sample is for
the NN training. It consists of 40k  $e^{+}e^{-}$  PYTHIA events
with the ISR included.
One half of this sample contains  fully inclusive events but 
without the $e^+e^-\to t \bar{t}\to 6\> jets$ process, while
another half contains the all hadronic decays of $t\overline{t}$.
The top masses were generated according to the Breit-Wigner distribution
with the nominal (peak) positions flatly distributed in the interval
$150 - 200$ GeV.
The momenta of the final-state particles were  smeared as discussed in
Sect.~\ref{sec:dr}. 
The  second set contains also 40k  events, half
of them are the fully hadronic top decays. This set was used to assess
the performance (generalization) of the learning, in order to avoid
over-fitting.

\subsection{Neural network architecture and the results}

The neural network selection was based on the Stuttgart Neural Network
Simulator \cite{snns} with the logistic activation function
$1/(1+\exp (-x))$.
A feed-forward architecture with one output node representing the
probability of the observation of the fully hadronic top decays was
used. For an input, we use nine nodes representing $1-T,$ $Mj$, the
rescaling value of the invariant mass, $M_2$, and six values of $Y_{i}$
representing the Durham-algorithm resolution parameters.
We investigated the effect of varying the number of hidden layers and nodes in
these layers. It was found that 
one hidden layer with seven nodes was sufficient 
to achieve the best selection performance.

\begin{figure}
\begin{center}
\includegraphics[scale=0.5]{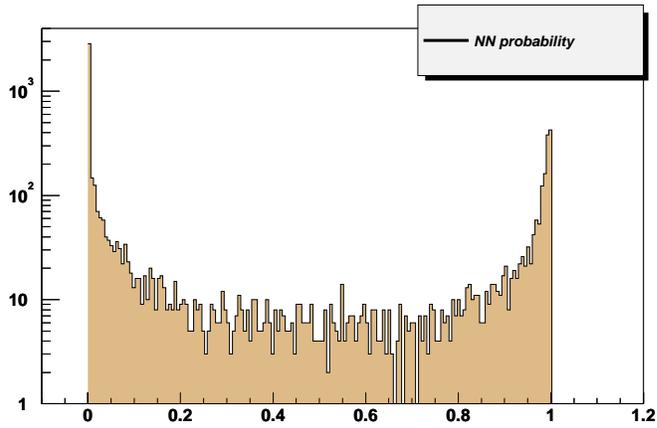}
\caption{The output of the neural network for the fully inclusive 
PYTHIA events after the Gaussian smearing of particle momenta to imitate
the LC detector response.}
\label{cap:NNout}
\end{center}
\end{figure}

The neural net was applied  to select the top events from the sample 
discussed in Sect.~\ref{sec:dr}.
Events were accepted if the NN output values are above 0.97. 
Figure~\ref{cap:NNout} shows the NN output, which has a clear peak near unity. 

Figure~\ref{cap:sim4} illustrates the reconstructed top events
after the NN event selection. 
The top reconstruction was the same as  for Fig.\ref{cap:sim1}. 
The signal-over-background ratio is 
larger than that without the NN selection, as well as the statistical error on the
top mass is smaller than for Fig.\ref{cap:sim2}-\ref{cap:sim3}. 
However, a small shift in the reconstructed mass was observed which
should properly be taken into account after a correction procedure. 
In addition, the shape of the background is less
understood than in case without the NN selection. 

%%%%%%%%%%%%%%%%%%%%%%%%%%%%%%%%%%%%%%%%%%%%%%%%%%
\section{Accuracy on the top-quark measurements}
%%%%%%%%%%%%%%%%%%%%%%%%%%%%%%%%%%%%%%%%%%%%%%%%%

The Monte Carlo luminosity used in this analysis corresponds to 
$16$ fb$^{-1}$ for $\sqrt{s}=500$ GeV and 33 fb$^{-1}$ for 
$\sqrt{s}=800$ GeV.
Our results indicate that the statistical
error on the top-mass measurement for this luminosity is about $380-450$ MeV. Thus,
for an integrated luminosity of 300  fb$^{-1}$, corresponding to one to two
years of running, 
the proposed method for the all-hadronic decay channel 
will lead to a statistical error of 
$\delta m_t \sim 100$ MeV at $\sqrt{s}=500$ GeV.
The statistical uncertainty on the mass measurement for
$\sqrt{s}=800$ GeV is by a factor two larger due to
a significant missing energy from neutrinos. 
Tables~\ref{tab2} and \ref{tab3} show the  results for $\sqrt{s}=500$ GeV and $\sqrt{s}=800$ GeV.

The direct measurement of the top-quark width in this decay mode is difficult.
The reconstructed width is by more than a factor ten larger than the
generated (true) top width.
Therefore, other methods should be investigated to be able to measure the top width from the
hadronic-final-state signatures of $e^+e^-$ collisions. 

One can also roughly estimate the statistical accuracy on the measurements
of the total cross section for the all-hadronic top-decay channel.
This study indicated that the number of the top candidates is around
2k for 16 fb$^{-1}$.
This number of the top candidates was obtained from 4224 $t\bar{t}\to
6\> jets$ events, thus
the acceptance is $\sim 24\%$. For an integrated luminosity of 300  
fb$^{-1}$, about 80k $t\bar{t}\to 6\> jets$ events are expected
assuming $\sigma_{t\bar{t}}=0.6$ pb. This leads to  
$\sim 260 \pm 0.9$ fb for the observed cross section, thus the relative accuracy is
$0.35\%$.
After the event reconstruction, $\sim 19200 \pm 124$ events are expected, leading to
$64.8 \pm 0.4$ fb, and to a relative accuracy of $0.6\%$.
 
%%%%%%%%%%%%%%%%%%%%%
\section{Summary}
%%%%%%%%%%%%%%%%%%%%%

For the first time, the all-hadronic top decays were studied
using the full TESLA detector simulation, after taking into account a realistic
contribution from multi-jet QCD background. This was done using a new
method of the reconstruction, which is expected to be less sensitive to
experimental and theoretical uncertainties arising from the 
direct measurement of the dijets with the invariant masses close the
$W$ mass. Note that the method is a very general and can be used for
the reconstruction of any two particles decaying into six jets in $e^+e^-$ 
collisions. 
Also, the proposed approach is significantly simpler than the  
top-reconstruction method adopted at the TEVATRON, where a 
kinematic fit is used.  
The present method takes advantage of the fact that,
in $e^+e^-$ annihilation, the entire
hadronic event can be  reconstructed. 
Therefore, essentially all produced particles can be grouped
into jets using an exclusive jet algorithm, while 
energy-momentum conservation provides an additional
handle for the top reconstruction.
In contrast, the reconstruction of top quarks 
at the TEVATRON and LHC is characterized by a large missing
energy along the beam directions, and  by the use of 
inclusive jet algorithms which combine 
only a relatively small fraction of the produced hadrons into jets.  
We did not use the double $b$-tagging to obtain the top signal; 
we have verified that the $b$-tagging reduces the QCD background, but does not have 
a significant impact on the reconstructed width of the top quark.

The TESLA detector resolution
for the top-mass measurement based on the energy-flow algorithm is
$5-5.5$ GeV for $\sqrt{s}=500$ GeV. This value is almost independent of 
method used to reconstruct the top quarks. For  $\sqrt{s}=800$ GeV,
the detector resolution is $9.5$ GeV. Such an increase is due to 
higher jet energies, larger overlaps between energy deposits used for
the energy-flow objects, as well as a larger calorimeter  
leakage. These two numbers can be used for comparisons with 
other detector designs and/or
other methods of the event reconstruction.   
Note that even before the detector simulation, a significant 
width for the three-jet mass spectrum due to fragmentation effects
is observed, thus the reconstruction of the top width from hadronic jets 
is very unlikely.  
     
In this paper, we have estimated the typical statistical uncertainties  on the
top mass measurement, as well as the statistical uncertainty
anticipated  for the top-quark production cross section in the all-hadronic
decay channel. 
The method leads to a statistical uncertainty of $\delta m_t \simeq
100$ MeV for the modest
value of the luminosity, 300 fb$^{-1}$. 
This uncertainty is compatible with the
statistical precision for the most 
promising lepton-plus-jets decay channel at the LHC \cite{lhc}, and
is well below the theoretical systematic uncertainty anticipated for the pole top mass.
An essential aspect of this method is to understand limitations arising from 
experimental systematic uncertainties. However, this study has not been carried out yet.

%%%%%%%%%%%%%%%%%%%%%%%%% include references here
\newpage
\providecommand{\etal}{et al.\xspace}
\providecommand{\coll}{Collab.\xspace}
\catcode`\@=11
\def\@bibitem#1{%
\ifmc@bstsupport
  \mc@iftail{#1}%
    {;\newline\ignorespaces}%
    {\ifmc@first\else.\fi\orig@bibitem{#1}}
  \mc@firstfalse
\else
  \mc@iftail{#1}%
    {\ignorespaces}%
    {\orig@bibitem{#1}}%
\fi}%
\catcode`\@=12

\newpage

\begin{table}[tbp]
\begin{tabular}{|l|c|c|c|c|c|c|}
\hline     & $\Delta_E$  & $\Delta_{PL}$  & $\Delta_{PT}$ & $\Delta_{y}$  &  $\Delta_{M}$ (GeV) & $\Delta_{P}$ (GeV)  \\ 
% \hline Generator level & 0.02 & 0.02 & 0.02 & $2\cdot 10^{-4}$ & 30 & 20   \\ 
\hline a)  PYTHIA, $\sqrt{s}=500$ GeV & 0.07 & 0.04  & 0.04 & $2\cdot 10^{-4}$ & 40  & 20  \\ 
\hline b)  PYTHIA, $\sqrt{s}=800$ GeV & 0.4 & -  & 0.3 & $2\cdot 10^{-4}$ & 40  & 20  \\
\hline 
\end{tabular} 
\caption{The parameters used for the selection and 
reconstruction of the top quarks in $e^+e^-$ annihilation at: 
a) $\sqrt{s}=500$ GeV; b) $\sqrt{s}=800$ GeV.}
\label{tab1}
\end{table}

\begin{table}[tbp]
\begin{tabular}{|l|c|c|c|c|c|c|}
\hline   PYTHIA, $\sqrt{s}=500$ GeV  & $\delta m_t$ (GeV)  & BRW width (GeV)  & Gaussian width (GeV) \\ 
\hline Generated           &  0.030  & $9.5 \pm 0.4$ & -   \\ 
\hline a) Reconstructed       &  0.105  & 9.5 (fixed) & $5.0\pm 0.2$    \\ 
\hline b) Reconstructed  (NN) &  0.087  & 9.5 (fixed) & $5.2\pm 0.1$  \\ 
\hline 
\end{tabular} 
\caption{The statistical uncertainties on the top-mass measurement assuming an integrated
luminosity of $300$ fb$^{-1}$ for $e^+e^-$ annihilation events at $\sqrt{s}=500$ GeV. 
The table also shows the typical reconstructed widths 
for the cases:  a) without using the NN 
(see Fig.~\ref{cap:sim2}) and b) with the NN selection 
(see Fig.~\ref{cap:sim4}).}
\label{tab2}
\end{table}

\begin{table}[tbp]
\begin{tabular}{|l|c|c|c|c|c|c|}
\hline   PYTHIA, $\sqrt{s}=800$ GeV  & $\delta m_t$ (GeV)  & BRW width (GeV)  & Gaussian width (GeV) \\ 
\hline Generator level & 0.04 & $7.7\pm 0.4$ & -   \\ 
\hline Reconstructed level &  0.230 & 7.7 (fixed) & $9.5\pm 0.4$   \\ 
\hline 
\end{tabular} 
\caption{
The statistical uncertainties on the top-mass measurement for $e^+e^-$ annihilation events 
at $\sqrt{s}=800$ GeV assuming an integrated 
luminosity of $300$ fb$^{-1}$. The table also shows the 
typical reconstructed widths.} 
\label{tab3}
\end{table} 

\newpage
%%%%%%%%%%%%%%%%%%%%%%%%%%%%%%%%%%%%%%%%%%% figures

\begin{figure}
\begin{center}
\includegraphics[  scale=0.5]{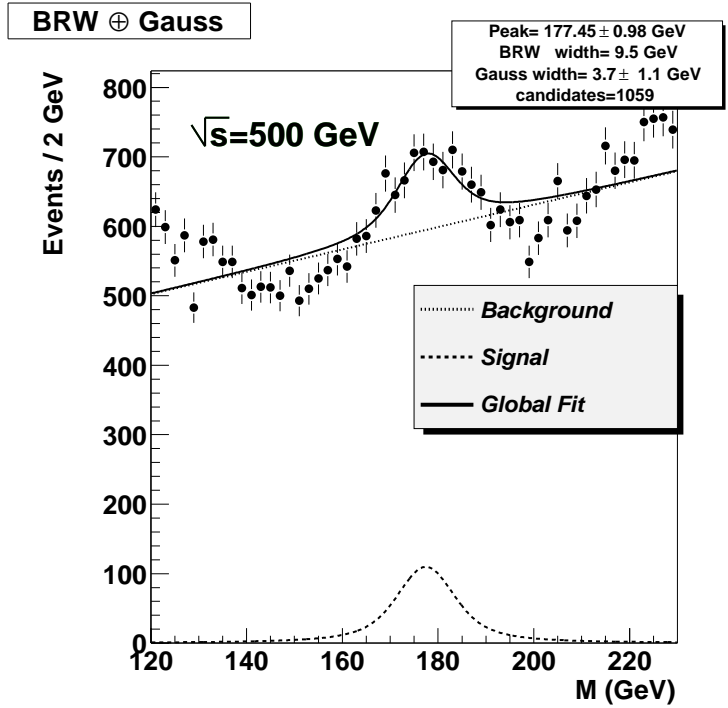}
\caption{
The invariant-mass distributions for three-jet pairs obtained from
the fully inclusive $e^+e^-$ PYTHIA events. 
Events $e^+e^- \to t\bar{t}$
were passed through the full detector simulation, while the continuum
was obtained after the Gaussian smearing of 
the original particle momenta.
The top events 
were selected and reconstructed using the parameters 
given in Table~\ref{tab1}a), but without the cut on $y_6^{cut}$.
The fit is based on  the Breit-Wigner function convoluted with a
Gaussian, plus a liner function to describe the background. The width
of the Breit-Wigner distribution is fixed to 9.5 GeV (see
Fig.~\ref{cap:tru1}).  
}
\label{cap:sim1}
\end{center}
\end{figure}

\begin{figure}
\begin{center}
\includegraphics[  scale=0.5]{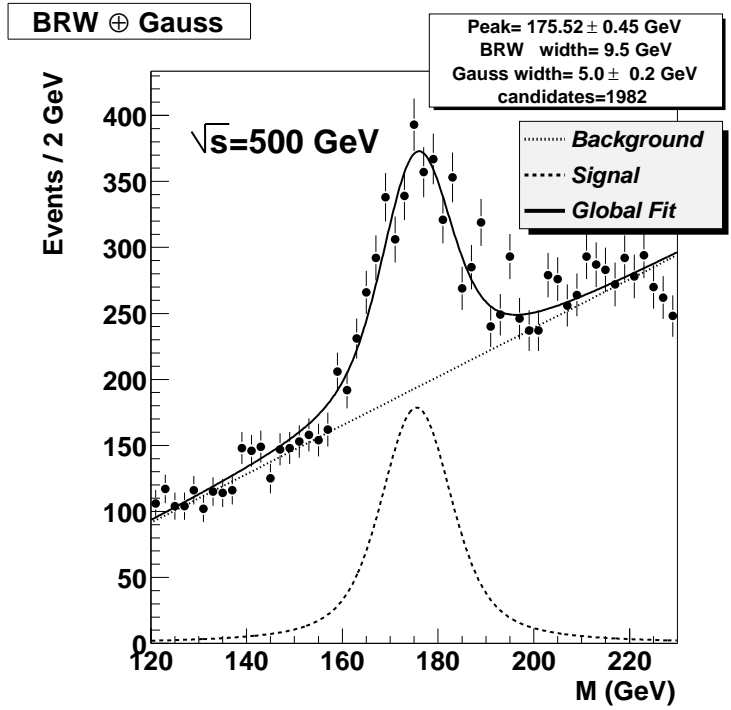}
\caption{
Same as Fig.~\ref{cap:sim1}, but after applying the cut 
$y_6^{cut}>2\cdot 10^{-4}$.}
\label{cap:sim2}
\end{center}
\end{figure}

\begin{figure}
\begin{center}
\includegraphics[  scale=0.5]{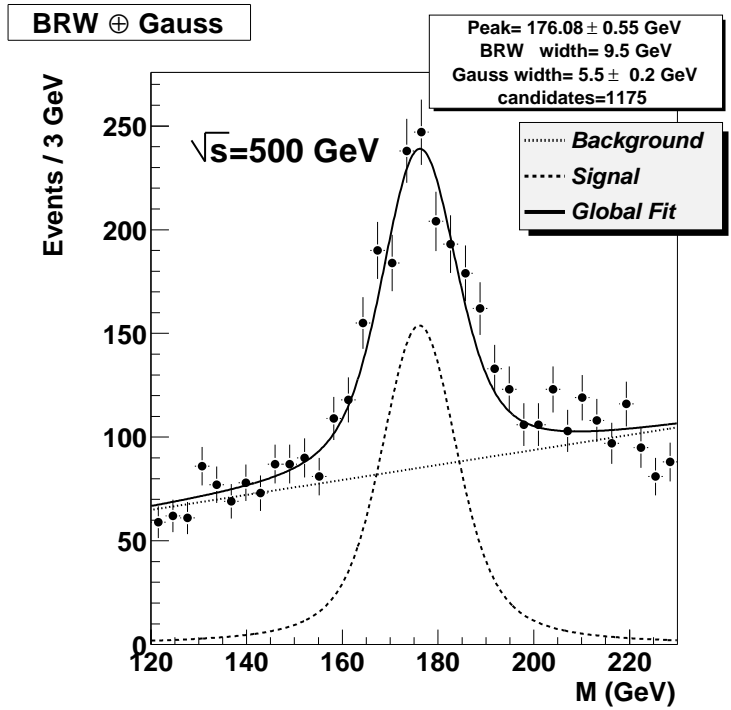}
\caption{
Same as Fig.~\ref{cap:sim2}, but in addition, only three-jet
groups  are shown which have at least one jet pair  
with  $\mid M_{jj}- M_W\mid <15$ GeV.}
\label{cap:sim3}
\end{center}
\end{figure}

\begin{figure}
\begin{center}
\includegraphics[  scale=0.55]{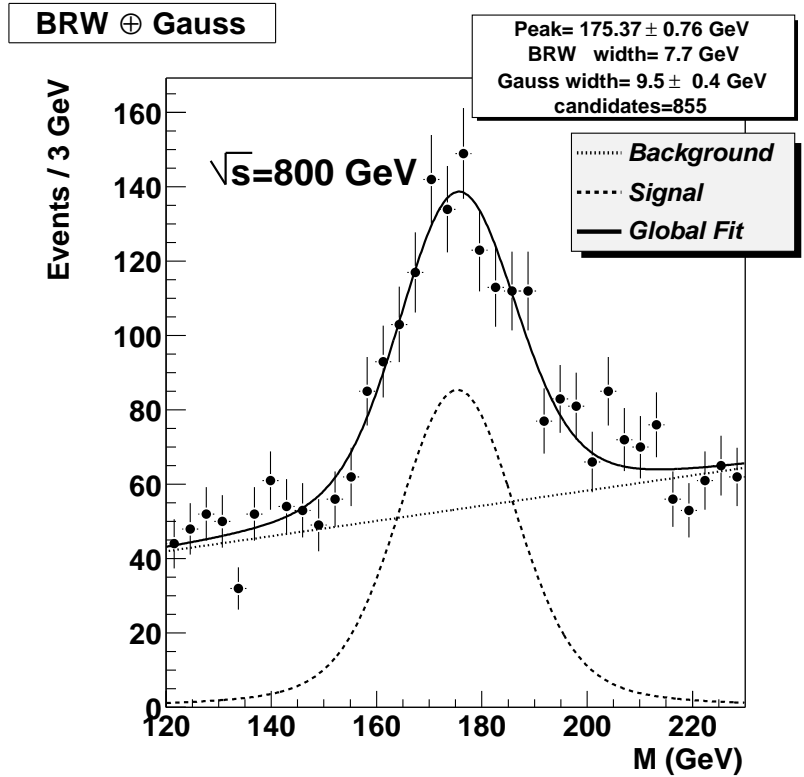}
\caption{The reconstructed invariant masses for the all-hadronic top-decay
channel generated with PYTHIA for $\sqrt{s}=800$ GeV after the LC detector simulation.
The applied cuts as for Fig.~\ref{cap:sim2}. The Breit-Wigner width is
taken from Fig.~{cap:tru2}.}
\label{cap:sim5}
\end{center}
\end{figure}

\newpage

\newpage
\begin{figure}
\begin{center}
\includegraphics[  scale=0.5]{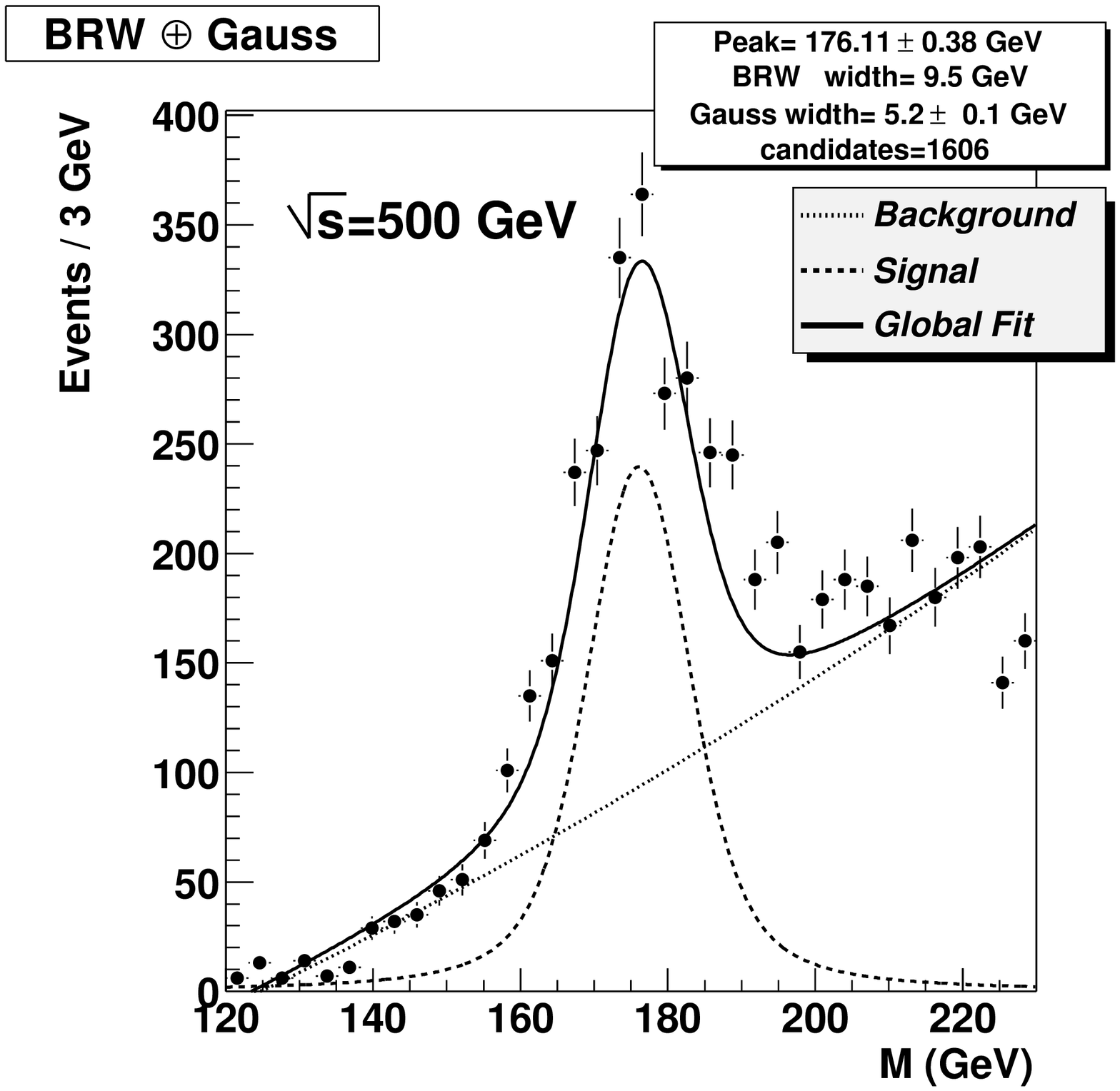}
\caption{
The invariant-mass distribution for three-jet clusters after the NN selection.
The event selection as for Fig.~\ref{cap:sim1}.
}
\label{cap:sim4}
\end{center}
\end{figure}


\begin{mcbibliography}{10}

\bibitem{tevat}
K.~Tollefson and E.W.~Varnes,
\newblock Annu. Rev. Nucl.{} {\bf 49},~435~(1999)\relax
\relax
\bibitem{lhc}
{\em Top quark physics}, ed.~M.~Beneke et al.
\newblock Proceedings of the workshop "1999 CERN Workshop on SM physics at the
  LHC", {\tt hep-ph/0003033}, 1999\relax
\relax
\bibitem{iwasaki}
M.~Iwasaki.
\newblock Presented at the world-wide study of physics and detectors for future
  linear colliders (LCWS99), (Sitges, Barcelona, Spain, 1999) OREXP 99-04, {\tt
  hep-ex/9910065} (1999)\relax
\relax
\bibitem{epj:c9:229}
S.~Moretti,
\newblock \EPJ{} {\bf C~9},~229~(1999)\relax
\relax
\bibitem{np:b544:289}
S.~Moretti,
\newblock \NP{} {\bf B~544},~289~(1999)\relax
\relax
\bibitem{gangemi}
F.~Gangemi et al.,
\newblock \NP{} {\bf B~559},~3~(1999)\relax
\relax
\bibitem{che:lcsim}
S.V.~Chekanov,
\newblock \EPJ{} {\bf C~26},~173~(2002)\relax
\relax
\bibitem{pl:b434:115}
M.~Beneke,
\newblock \PL{} {\bf B~434},~115~(1998)\relax
\relax
\bibitem{pr:d59:114014}
A.H.~Hoang et al.,
\newblock \PR{} {\bf D~59},~114014~(1999)\relax
\relax
\bibitem{EPJdirect:c3:1}
A.H.~Hoang et al.,
\newblock EPJdirect{} {\bf C~3},~1~(2000)\relax
\relax
\bibitem{zp:c62:281}
T.~Sj{\"o}strand and V.A.~Khoze,
\newblock \ZP{} {\bf C~62},~281~(1994)\relax
\relax
\bibitem{epj:c6:271}
V.A.~Khoze and T.~Sj{\"o}strand,
\newblock \EPJ{} {\bf C~6},~271~(1999)\relax
\relax
\bibitem{zp:c72:71}
A.~Ballestrero et al.,
\newblock \ZP{} {\bf C~72},~71~(1996)\relax
\relax
\bibitem{epjdir:c1:1}
V.A.~Khoze and T.~Sj{\"o}strand,
\newblock Eur. Phys. J. direct{} {\bf C~2},~1~(2000)\relax
\relax
\bibitem{kh_sj_top}
V.A.~Khoze and T.~Sj{\"o}strand,
\newblock {\em {QCD} {I}nterconnection {S}tudies at {L}inear {C}ollider},
\newblock in {\em Physics and Experimentation at a Linear Electron-Positron
  Collider, Vol.1}, ed.~T.~Behnke et al., p.~257.
\newblock DESY-01-123F, 2001\relax
\relax
\bibitem{pl:B351:293}
L.~L{\"o}nnblad and T.~Sj{\"o}strand,
\newblock \PL{} {\bf B~351},~293~(1995)\relax
\relax
\bibitem{epj:c6:403}
S.V.~Chekanov, E.A.~De Wolf, W.~Kittel,
\newblock \EPJ{} {\bf C~6},~403~(1999)\relax
\relax
\bibitem{pl:b269:432}
S.~Catani et al.,
\newblock \PL{} {\bf B~269},~432~(1991)\relax
\relax
\bibitem{cpc:82:7}
T.~Sj{\"o}strand et al.,
\newblock \CPC{} {\bf 135},~238~(2001)\relax
\relax
\bibitem{tdr-tesla}
{\em Technical Design Report, Part IV, "A Detector for TESLA"}, ed.~T.~Behnke
  et al.
\newblock DESY-2001-011, 2001\relax
\relax
\bibitem{brahms}
T.~Behnke \etal,
\newblock {\em {BRAHMS:} {A} {M}onte {C}arlo for a {D}etector at 500/800~\gev
  ~{L}inear {C}ollider},
\newblock in {\em Physics and Experimentation at a Linear Electron-Positron
  Collider}, ed.~T.~Behnke \etal, p.~1425.
\newblock DESY-01-123F, {\tt
  http://www-zeuthen.desy.de/lc\_repository/detector\_simulation/dev/BRAHMS/},
  2001\relax
\relax
\bibitem{morgunov:2001cd}
V.L.~Morgunov,
\newblock {\em {E}nergy-flow {M}ethod for {M}ulti-jet {E}ffective mass
  {R}econstruction in the {H}ighly {G}ranulated {TESLA} {C}alorimeter}.
\newblock APS/DPF/DPB Summer Study on the Future of Particle Physics (Snowmass
  2001), {\tt http://www.slac.stanford.edu/econf/C010630/proceedings.shtml},
  2001\relax
\relax
\bibitem{igo-1992}
P.~Igo-Kimens and J.~H.~K{\"u}hn,
\newblock {\em Proc. of the {W}orkshop on "$e^+e^-$ {C}ollisions at 500 gev.
  {T}he {P}hysics {P}otential"}.
\newblock (Munich-Annecy-Hamburg) DESY 92-123, 1992\relax
\relax
\bibitem{pl:12:54}
S.~Brandt et al.,
\newblock \PL{} {\bf 12},~54~(1964)\relax
\relax
\bibitem{prl:39:1587}
E.~Fahri,
\newblock \PRL{} {\bf 39},~1587~(1977)\relax
\relax
\bibitem{pl:b278:181}
L.~L{\"o}nnblad, C.~Peterson and T.R{\"o}gva,
\newblock \PL{} {\bf B~278},~181~(1992)\relax
\relax
\bibitem{snns}
A.~Zell et al.,
\newblock {\em Stuttgart Neural Network Simulator (SNNS),Version 4.2}.
\newblock Unpablished\relax
\relax
\end{mcbibliography}
\end{document}